# Comparison of 3D-DXA Measurements with QCT Software Solutions for Proximal Femur Bone Density Assessment

**Authors:** Asier Muñoz[1]*, Luis del Rio[2], Jorge Malouf[3], and Ludovic Humbert[1]

**Affiliations:**

[1]3D-Shaper Medical, Barcelona, Spain

[2]CETIR ASCIRES, Barcelona, Spain

[3]Mineral Metabolism Unit, Grup Creu Groga, Mataró, Spain

* Corresponding author:
Asier Muñoz
3D-Shaper Medical, Rambla Catalunya 53-55, 4th floor, 08007 Barcelona, Spain
Email: asier.munoz@3d-shaper.com

## Abstract

*3D-Shaper* reconstructs proximal femur shape and density in 3D from a standard hip dual-energy X-ray absorptiometry (DXA) scan. In men and women aged 35–91 years, we compared *3D-Shaper* integral volumetric bone mineral density (vBMD), trabecular vBMD, and cortical surface BMD (sBMD) with quantitative computed tomography (QCT) processed with three methods: an in-house pipeline, *Stradview* and *QCTPro CTXA*, which were used to derive bone measurements equivalent to those of *3D-Shaper*. Agreement, including among QCT methods, was assessed with linear regression and Bland-Altman analysis (bias and limits of agreement, LoA). *3D-Shaper* agreed strongly with the in-house pipeline and *Stradview* for integral vBMD (R = 0.95–0.96; bias 1.5 to 4.2 mg/cm$^3$; LoA ±39.7 to ±41.4 mg/cm$^3$), trabecular vBMD (R = 0.92–0.94; bias −11.8 to −7.8 mg/cm$^3$; LoA ±33.3 to ±39.4 mg/cm$^3$) and cortical sBMD (R = 0.94; bias 5.6 to 38.2 mg/cm$^2$; LoA ±23.6 to ±26.4 mg/cm$^2$). The in-house pipeline and *Stradview* agreed for integral vBMD (R = 1.00; bias −2.7 mg/cm$^3$; LoA ±3.7 mg/cm$^3$), trabecular vBMD (R = 0.96–1.00; bias −0.2 to 3.9 mg/cm$^3$; LoA ±6.9 to ±27.5 mg/cm$^3$) and cortical sBMD (R = 0.98; bias 32.7 mg/cm$^2$; LoA ±14.6 mg/cm$^2$). Compared with *3D-Shaper*, *QCTPro CTXA* integral vBMD was systematically lower (R = 0.90; bias −28.4 mg/cm$^3$; LoA ±61.5 mg/cm$^3$) and trabecular vBMD showed a proportional bias (R = 0.89; bias −44.1 mg/cm$^3$; LoA ±61.3 mg/cm$^3$). The same pattern was observed against the other QCT methods, for integral vBMD (R = 0.93; bias −32.6 to −29.9 mg/cm$^3$; LoA ±49.6 to ±50.0 mg/cm$^3$) and trabecular vBMD (R = 0.94–0.96; bias −36.3 to −32.5 mg/cm$^3$; LoA ±51.2 to ±54.9 mg/cm$^3$). These findings support the agreement of *3D-Shaper* with independent QCT approaches and show that *3D-Shaper*–QCT differences must be interpreted in light of the QCT method used.

## 1. Introduction

Osteoporotic hip fractures are a major driver of disability, mortality, and healthcare costs, with their incidence projected to rise alongside the aging global population [1,2]. Areal bone mineral density (aBMD) measured by dual-energy X-ray absorptiometry (DXA) remains the clinical standard for diagnosing osteoporosis and predicting fracture risk [3]. However, standard DXA-based screening exhibits a well-documented diagnostic limitation: the majority of fragility fractures occur in patients whose aBMD does not meet the World Health Organization (WHO) threshold for osteoporosis (T-score ≤ -2.5) [4–6]. This diagnostic gap reflects the multifactorial nature of fracture risk, but also stems in part from DXA's two-dimensional (2D) projection, which yields a single areal density measurement rather than a three-dimensional (3D) density distribution, and which cannot distinguish cortical from trabecular bone.

Quantitative computed tomography (QCT) overcomes this limitation by reconstructing bone in three dimensions, yielding true volumetric bone mineral density (vBMD) and enabling independent quantification of cortical and trabecular compartments [7]. Although QCT is used clinically at a number of sites and is increasingly applied opportunistically to CT scans acquired for other indications [8,9], it remains far less widely used than DXA, owing to higher radiation exposure and cost, and the need for specialized software and trained operators. Furthermore, unlike DXA, QCT lacks standardized acquisition and analysis protocols: scanner models, calibration phantoms, and reconstruction settings vary widely across sites, while different analysis software employ distinct segmentation algorithms that can yield systematically different estimates of cortical and trabecular bone from the same underlying scan [10], limiting the comparability of QCT-derived measurements across studies and centers.

Three-dimensional dual-energy X-ray absorptiometry (3D-DXA) has been proposed to bridge this gap between DXA and QCT [11]. A commercial implementation of this approach, *3D-Shaper* (3D-Shaper Medical, Barcelona, Spain), computes integral vBMD, trabecular vBMD, and cortical surface bone mineral density (sBMD) at the proximal femur. From a single DXA scan, the software reconstructs a patient-specific 3D model of the proximal femur through an iterative 2D-3D registration process, in which a statistical shape and density model (SSDM) built from a database of QCT scans is deformed until its projection matches the DXA image [11]. Because acquisition relies on a conventional DXA scan, 3D-DXA preserves the low radiation dose and clinical accessibility of DXA while providing compartment-specific information that DXA alone cannot resolve. *3D-Shaper* has been used to evaluate different treatment strategies for skeletal fragility, with particular attention to the differential effects of treatment on cortical and trabecular bone density [12–19].

Agreement between *3D-Shaper* and QCT has been assessed in multiple validation studies [11,18,20–22]. At the total proximal femur, reported correlation coefficients (R) have ranged from 0.84 to 0.96 for integral vBMD [11,18,22], from 0.79 to 0.95 for trabecular vBMD [11,18,22], and from 0.85 to 0.95 for cortical sBMD [22]. A recent multicenter evaluation across diverse international cohorts confirmed high accuracy and generalizability, exhibiting strong correlations ($R = 0.82 – 0.97$), with small biases ranging from −14.75 to 4.50 mg/cm$^3$ for integral vBMD, −8.31 to 14.41 mg/cm$^3$ for trabecular vBMD, and −5.58 to 3.21 mg/cm$^2$ for cortical sBMD, and limits of agreement (LoA) ranging from ±32.4 to ±39.0 mg/cm$^3$, ±26.5 to ±36.2 mg/cm$^3$, and ±17.9 to ±22.3 mg/cm², respectively [22]. In the STRUCTURE study, correlations with QCT were also strong for integral and trabecular vBMD (R = 0.84–0.87 and 0.79–0.86, respectively) [18]. An independent ex vivo evaluation using 77 femora reported a correlation (R = 0.96) and LoA (±41 mg/cm³) for integral vBMD at the total femur in line with previous studies, but a larger bias, with *3D-Shaper* values systematically lower than QCT by 64 mg/cm³ on average [21]. The authors attributed part of this difference to their ex vivo setup: femora were scanned between water bags to simulate soft tissue during DXA acquisition, although water is an adequate surrogate for lean but not for fat tissue, and a solid rather than liquid calibration phantom was used for QCT.

QCT-derived bone measurements are themselves method-dependent. When two established analysis packages, *QCTPro CTXA* (computed tomography X-ray absorptiometry; Mindways Software, Austin, TX, USA) and *MIAF-Femur* (University of Erlangen, Erlangen, Germany), were applied to the same CT scans, integral vBMD at the total femur agreed reasonably well (R = 0.95), with a bias of 32.7 mg/cm³, but trabecular and cortical density diverged considerably [10]. Trabecular vBMD also showed a strong correlation (R = 0.94), but with a larger bias of 57 mg/cm³, whereas cortical vBMD showed much poorer correlation (R = 0.33), with a bias of 324.6 mg/cm³ [10]. These findings, as the authors noted, underscore that QCT-derived measurements, particularly cortical and trabecular measurements, can depend substantially on the analysis method used to define the underlying bone compartments. Such methodological differences are relevant when interpreting comparisons between *3D-Shaper* and QCT. Among the available QCT approaches, *Stradview* (University of Cambridge, Cambridge, UK) provides a complementary model-based analysis of the proximal femur, estimating cortical thickness and density by fitting the local CT density profile across the cortex [23].

The primary objective of this study is to evaluate the agreement of *3D-Shaper* measurements derived from DXA with multiple QCT software approaches. We compared *3D-Shaper* parameters against three QCT-based methods: an in-house QCT pipeline previously used to validate *3D-Shaper* [11,22], *Stradview*, and *QCTPro CTXA*. Furthermore, we evaluated the agreement between these QCT-based methods themselves, thereby contextualizing any observed differences between *3D-Shaper* and QCT within the scope of the inherent variability across QCT methods.

## 2. Materials and Methods

### 2.1. Study Subjects

The study included 60 subjects with paired DXA and QCT scans acquired within a 2-month interval. This exact dataset was included in the original *3D-Shaper* validation study [11] and in a subsequent multicenter validation of *3D-Shaper* against QCT [22]. None of these subjects were included in the QCT database used to build the *3D-Shaper* SSDM. Subjects with skeletal diseases other than osteoporosis (e.g., osteoarthritis or Paget's disease) or a history of hip fracture were not included. The study was approved by the Institutional Review Board or ethics committee of the clinical center, and written informed consent was obtained from all study subjects.

### 2.2. DXA Acquisition and 3D-Shaper Analysis

DXA scans were acquired at CETIR Centre Mèdic (Barcelona, Spain) using a Lunar iDXA scanner (GE Healthcare, Madison, WI, USA), and total hip and femoral neck aBMD values were derived using enCORE software. Total hip and femoral neck aBMD T-scores were calculated using NHANES III normative data, following international guidelines [24]. The lower of the two T-scores was used to classify each subject into one of three categories: very low bone density (T-score ≤ −2.5), low bone density (−2.5 < T-score < −1), or normal bone density (T-score > −1).

Proximal femur 3D-DXA analysis was carried out with *3D-Shaper* software, version 2.14 (3D-Shaper Medical, Barcelona, Spain). *3D-Shaper* reconstructs a patient-specific model of the proximal femur from a single anteroposterior DXA image using an SSDM built from a database of QCT scans independent of the present study population [11]. The SSDM captures the principal modes of variation in both femoral shape and bone density observed across the training database. The shape and density parameters of the SSDM, together with pose and scale, are optimized iteratively: at each iteration, a subject-specific instance of the model is generated, projected in the anteroposterior direction, and compared against the actual DXA image, with the projection's similarity to the DXA image used to guide the optimization. This process yields a subject-specific 3D reconstruction of the proximal femur, comparable to a QCT image, from which the cortical and trabecular compartments are subsequently separated using a model-based segmentation approach [25]. At each vertex of the reconstructed femoral surface, density variations across the cortex are modeled as a function of the cortical thickness, the density of the cortex, the density of surrounding tissue and trabecular bone, and the blurring introduced by the imaging system, with these parameters estimated by fitting the model to the interpolated density profile, measured along a line normal to the surface, using the Levenberg-Marquardt algorithm. Because thin cortices, relative to the imaging resolution, make thickness and density difficult to resolve independently, the fitting is constrained using a prior relationship between cortical thickness and density, derived from high-resolution micro-CT scans of cadaveric femurs [25]. In its current clinical release (version 2.14), *3D-Shaper* outputs three parameters at the total femur: integral vBMD ($mg/cm^3$) (*3D-Shaper Int.vBMD*), trabecular vBMD ($mg/cm^3$) (*3D-Shaper Tb.vBMD*), and cortical sBMD ($mg/cm^2$) (*3D-Shaper Ct.sBMD*), the product of cortical thickness and cortical vBMD, reflecting the mass of cortical bone per unit surface area rather than per unit volume.

### 2.3. QCT Acquisition and Analysis

QCT scans were acquired at two centers: CETIR Esplugues PET (Esplugues de Llobregat, Spain), using a Philips Gemini GXL 16 scanner (Philips Healthcare, Best, The Netherlands; n = 39), and CETIR Clínica Pilar Sant Jordi RM (Barcelona, Spain), using either a GE Discovery CT750 HD scanner (n = 10) or a GE HiSpeed QX/i scanner (GE Healthcare, Chicago, IL, USA; n = 11). At both centers, a liquid $K_2HPO_4$-equivalent calibration phantom (Model 3, Mindways Software, Austin, TX, USA) was scanned alongside each subject to be able to convert Hounsfield units (HU) to bone density values. Image acquisition parameters were as follows: 512 × 512-pixel matrix with in-plane pixel dimensions between 0.7 × 0.7 mm and 1.1 × 1.1 mm (mean, 0.9 × 0.9 mm); 0.5 mm inter-slice spacing; slice thickness of 1.0–1.25 mm; reconstruction with kernel A on the Philips unit and the standard kernel on the GE Healthcare units; and exposure settings of 120 kVp with tube current between 138 and 250 mA (110–150 effective mAs where available). CTDIvol, available for the Philips scanner, ranged from 10.6 to 11.6 mGy.

Three QCT-based methods were used to analyze the acquired scans: an in-house QCT pipeline previously used to validate *3D-Shaper* [11,22], *Stradview* (University of Cambridge, Cambridge, UK), and *QCTPro CTXA* (Mindways Software, Austin, TX, USA). Each method was applied independently to the same QCT scans, using its own segmentation and calibration approach, and each provided values for the total femur region of interest matching the *3D-Shaper* parameters described above. The three methods, and how their outputs were matched to the *3D-Shaper* parameters, are described in turn below.

### 2.3.1. In-House QCT Pipeline

The proximal femur ipsilateral to the DXA acquisition was segmented from the surrounding volume in MATLAB (MathWorks, Natick, MA, USA) using a semi-automatic thresholding approach [11]. Briefly, a sphere was manually positioned over the femoral head, and a threshold was applied to create a mask isolating bone from surrounding soft tissue; the sphere was then used to mask the femoral head, isolating the proximal femur without the femoral head. A second threshold was applied within the sphere to segment the femoral head separately, and the resulting mask was added to the proximal femur segmentation. A 3D closing followed by a filling operation was applied to obtain the final mask, which was visualized and, where necessary, manually corrected. Cortical and trabecular compartments were segmented using the same model and measurement conditions applied to *3D-Shaper* (Section 2.2) [25]. To ensure QCT and *3D-Shaper* parameters were computed over a matching anatomical region, the *3D-Shaper* femoral surface was rigidly registered onto the corresponding QCT femoral surface using the Coherent Point Drift algorithm. The total femur region of interest defined on the *3D-Shaper* model was then projected onto the registered QCT surface. Integral vBMD (*In-house Int.vBMD*), trabecular vBMD (*In-house Tb.vBMD*), and cortical sBMD (*In-house Ct.sBMD*) were computed on the QCT data within this region of interest, using the same definitions applied to *3D-Shaper*.

### 2.3.2. Stradview

*Stradview* (University of Cambridge, Cambridge, UK) estimates cortical thickness and density using the same type of model-based line-profile fitting applied to *3D-Shaper* and the in-house pipeline (Section 2.3.1): at each point on the femoral surface, a parametric model of the density profile across the cortex is fitted to the measured CT data along a line normal to the surface, as a function of cortical thickness, cortical density, the densities of the surrounding soft tissue and adjacent trabecular bone, and the imaging blur [23]. The two implementations differ in how this ambiguity is resolved when the cortex is thin relative to the imaging resolution. *Stradview*'s CBM v2 algorithm proceeds in two stages: a constant cortical density is first estimated for the whole bone, then refined locally at each point by comparing the imaging blur required to fit the model there against a global estimate of the scanner's true imaging blur, with the discrepancy indicating whether the initial density assumption was too high or too low. A more detailed technical description of the algorithm is available in the original methodological paper by Treece and colleagues [23].

The femoral surface used for the *Stradview* analysis was imported from the in-house QCT segmentation (Section 2.3.1). Line length was set to 18 mm, approximately three times the maximal expected cortical thickness. Mesh resolution was set to its highest value to maximize accuracy, as this did not substantially increase computational cost. Surface smoothing was kept minimal, since the segmented surface was visually clean and free of noise; the lowest smoothing level sufficient to remove any residual surface noise was therefore applied.

Output data for all subjects were exported and postprocessed in MATLAB (MathWorks, Natick, MA, USA). Three output parameters, calculated at each vertex of the femoral surface, were used for each subject: cortical thickness, cortical density, and trabecular density. Cortical and trabecular density values, exported in HU, were converted to density using the calibration phantom scanned alongside each subject (Section 2.3) and the calibration equation calculated by *Stradview*. Cortical sBMD (*Stradview Ct.sBMD*) was then calculated at each vertex as the product of cortical thickness and the calibrated cortical density. All local values were averaged over the total femur region of interest (ROI), which corresponded to the in-house QCT pipeline's ROI (Section 2.3.1). The parameter labelled by *Stradview* as trabecular density corresponds to the estimated trabecular density close to the cortex, sampled along the same perpendicular line used to derive cortical thickness and density; because this differs from the trabecular vBMD calculated by *3D-Shaper* and the in-house QCT pipeline, which is not restricted to the near-cortical region, we refer to this *Stradview* parameter as subcortical density (*Stradview Sc.vBMD*) throughout this manuscript.

To obtain a trabecular vBMD estimate more directly comparable to the *3D-Shaper* approach, the periosteal and endocortical surfaces produced by *Stradview* were exported and run through the same in-house QCT pipeline (Section 2.3.1). For each subject, integral vBMD (*Stradview Int.vBMD*) was calculated as the mean density within the periosteal envelope, spanning both cortical and trabecular bone, and trabecular vBMD (*Stradview Tb.vBMD*) as the mean density within the endocortical envelope. These parameters, while defined by *Stradview*'s own segmentation outlines, are comparable to the corresponding trabecular and integral vBMD values calculated by *3D-Shaper* and the in-house QCT pipeline (Section 2.3.1).

### 2.3.3. QCTPro CTXA

The commercial *QCTPro CTXA* module (Mindways Software, Austin, TX, USA) employs a multistep segmentation procedure in which the femur is initially delineated from surrounding soft tissue using a fixed attenuation threshold (default, 100 mg/cm$^3$), followed by local refinement of the periosteal boundary using a range parameter (default, −250 mg/cm$^3$). The resulting 3D femoral segmentation is projected into a simulated 2D DXA image after carefully aligning the segmented femur in a predefined coordinate system, and is partitioned according to the three standard DXA hip regions of interest: femoral neck, trochanter, and intertrochanter. These regions are subsequently mapped back onto the original 3D CT dataset to define corresponding volumes of interest, from which volumetric BMD (vBMD) is determined [26]. Total hip integral vBMD and volume, considered here as the total femur equivalent for comparison with the other QCT-based approaches and with *3D-Shaper*, were calculated by summing the corresponding measurements across the three regions after appropriate calibration with the phantom scanned synchronously. Within each region, cortical and trabecular bone were differentiated using a cortical threshold (Tc), which was maintained at the default value of 350 mg/cm$^3$ [10].

The finite spatial resolution of CT introduces a partial-volume (PV) effect at the periosteal boundary, where voxels contain a mixture of mineralized bone and surrounding soft tissue. Consequently, the apparent segmented bone volume is systematically overestimated relative to the underlying anatomical bone volume. *QCTPro CTXA* corrects for this effect by estimating an excess volume ($\Delta V_{pv}$) from the periosteal surface area and the spatial resolution of the CT scanner. This volume is subtracted from the segmented bone volume, while the measured bone mineral mass remains unchanged. Consequently, PV correction reduces the measured bone volume and increases the corresponding vBMD [27].

For compartment-specific measurements, *QCTPro CTXA* allows $\Delta V_{pv}$ to be assigned either to the cortical or trabecular compartment. In the default analysis, $\Delta V_{pv}$ is subtracted from cortical volume, and trabecular vBMD remains uncorrected. In the alternative analysis, $\Delta V_{pv}$ is instead subtracted from trabecular volume, resulting in corrected trabecular vBMD. The two analyses therefore use the same PV volume correction but apply it to different compartments. Because the total corrected bone volume is unchanged, both analyses yield the same corrected integral vBMD [27].

The uncorrected measurements (*CTXA (Raw) Int.vBMD; CTXA (Raw) Tb.vBMD*) are used as the primary measurements. PV-corrected measurements are additionally reported for comparison (*CTXA (Cor.) Int.vBMD; CTXA (Cor.) Tb.vBMD*). Corrected trabecular vBMD is obtained from the alternative analysis, and corrected integral vBMD is identical between the two analyses. Because *QCTPro CTXA* does not compute cortical thickness at the total femur, its output cannot be used to calculate cortical sBMD, and this parameter is therefore not reported for this pipeline.

### 2.4.Statistical Analysis

Continuous variables are reported as mean (standard deviation) and range, and categorical variables as counts. Agreement between pipelines was assessed for each parameter using linear regression and Bland-Altman analysis. For each pairwise comparison, the slope, intercept, Pearson correlation coefficient (R), and its p-value are reported, together with the Bland-Altman bias (mean difference) and 95% LoA (calculated as bias ± 1.96 SD). The significance of the bias was tested with a paired t-test. Statistical analyses were performed in MATLAB. A p-value $< 0.05$ was considered statistically significant.

## 3. Results

### 3.1.Subject Characteristics

Demographic and DXA-derived characteristics of the cohort are summarized in Table 1. The cohort had a mean age of 61.5 (SD 13.6) years and a mean BMI of 26.3 (SD 4.9) kg/m². Mean aBMD was 0.891 (SD 0.199) g/cm² at the total hip and 0.824 (SD 0.162) g/cm² at the femoral neck. Based on the lower of the total hip and femoral neck T-score, 20 subjects had very low bone density, 20 had low bone density, and 20 had normal bone density.

**Table 1.** Demographic and DXA-derived characteristics of the study cohort.

| **Characteristic** | **N = 60** |
|---|---|
| Age (years) | 61.5 (13.6) [35–91] |
| Sex, female/male | 38 / 22 |
| Femur side, left/right | 60 / 0 |
| BMI (kg/m²) | 26.3 (4.9) [16.7–39.9] |
| aBMD total hip (g/cm²) | 0.891 (0.199) [0.448–1.413] |
| aBMD femoral neck (g/cm²) | 0.824 (0.162) [0.504–1.388] |
| Bone density (very low/low/normal) | 20 / 20 / 20 |

*Data are mean (SD) [range] for continuous variables and counts for categorical variables.*

### 3.2.Comparative Analysis of Bone Density Estimates

Agreement between *3D-Shaper* and the in-house QCT pipeline, *Stradview*, and *QCTPro CTXA*, as well as among these three QCT-based methods themselves, was assessed for each parameter using linear regression and Bland-Altman analysis, presented in the order integral vBMD, trabecular vBMD, and cortical sBMD.

#### 3.2.1. Integral vBMD

*In-house Int.vBMD* and *Stradview Int.vBMD* both agreed very closely with *3D-Shaper Int.vBMD*, with slopes close to unity (0.996 and 0.995) and strong correlations (R = 0.96 and R = 0.95) (Figure 1). Bias and LoA were 4.2 mg/cm$^3$ (LoA ±39.7 mg/cm$^3$) for *In-house Int.vBMD* and 1.5 mg/cm$^3$ (LoA ±41.4 mg/cm$^3$) for *Stradview Int.vBMD*. *QCTPro CTXA* also agreed closely with *3D-Shaper Int.vBMD*, both raw (slope 1.022, R = 0.90; bias −28.4 mg/cm$^3$, LoA ±61.5 mg/cm$^3$) and corrected (slope 0.938, R = 0.90; bias −7.8 mg/cm$^3$, LoA ±61.5 mg/cm$^3$), with correction reducing the bias without changing the strength of the correlation. The LoA were also noticeably wider for *QCTPro CTXA* than for In-house or *Stradview* (±61.5 mg/cm$^3$ versus roughly ±40 mg/cm$^3$ for the other two).

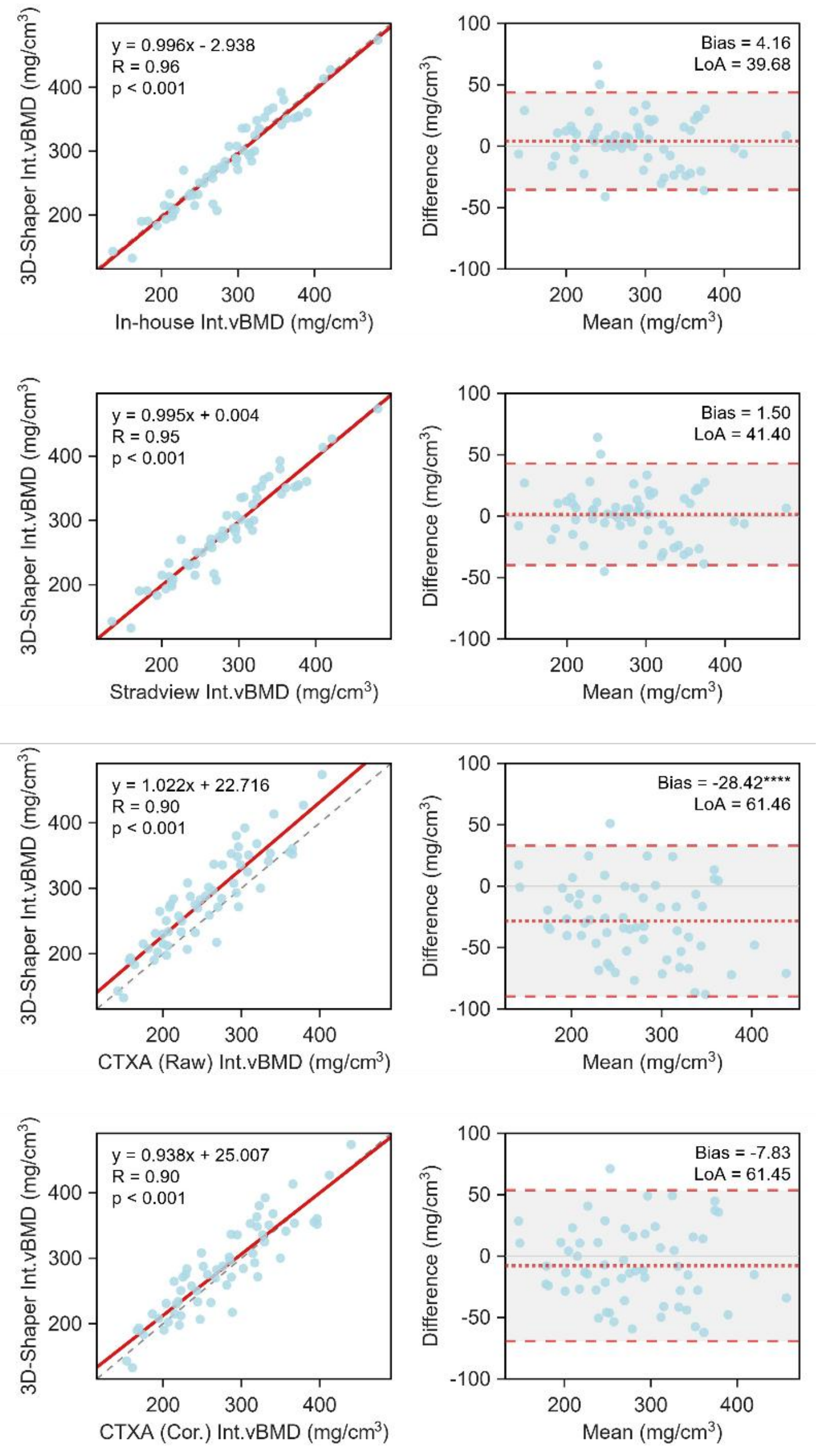

**Figure 1.** Agreement between *3D-Shaper Int.vBMD* and QCT-based methods, from top to bottom: *In-house Int.vBMD*, *Stradview Int.vBMD*, *CTXA (Raw) Int.vBMD*, and *CTXA (Cor.) Int.vBMD*. Left column: scatter plots, with the solid red line showing the linear regression fit (equation, Pearson's R, and p-value given in each panel) and the dashed gray line showing the identity line (y = x) for reference. Right column: corresponding Bland-Altman plots, showing the difference between the two measurements (x-axis minus y-axis method of the scatter plot) against their mean; the dotted red line marks the mean bias, the dashed red lines mark the 95% LoA (bias ± 1.96 SD), and the shaded band spans the LoA. Asterisks indicate the significance of the bias (paired t-test): * $p < 0.05$, ** $p < 0.01$, *** $p < 0.005$, **** $p < 0.001$.

Among the QCT-based methods, *Stradview Int.vBMD* agreed almost perfectly with *In-house Int.vBMD*, with a slope essentially equal to unity (1.003), a near-perfect correlation (R = 1.00), a small bias of −2.7 mg/cm³ (LoA ±3.7 mg/cm³), and no visible trend in the Bland-Altman differences (Figure 2). *CTXA (Raw) Int.vBMD* showed a weaker, more scattered agreement with both *In-house Int.vBMD* (slope 1.018, R = 0.93; bias −32.6 mg/cm³, LoA ±50.0 mg/cm³) and *Stradview Int.vBMD* (slope 1.016, R = 0.93; bias −29.9 mg/cm³, LoA ±49.6 mg/cm³), with a consistent negative bias but no clear proportional trend across the density range.

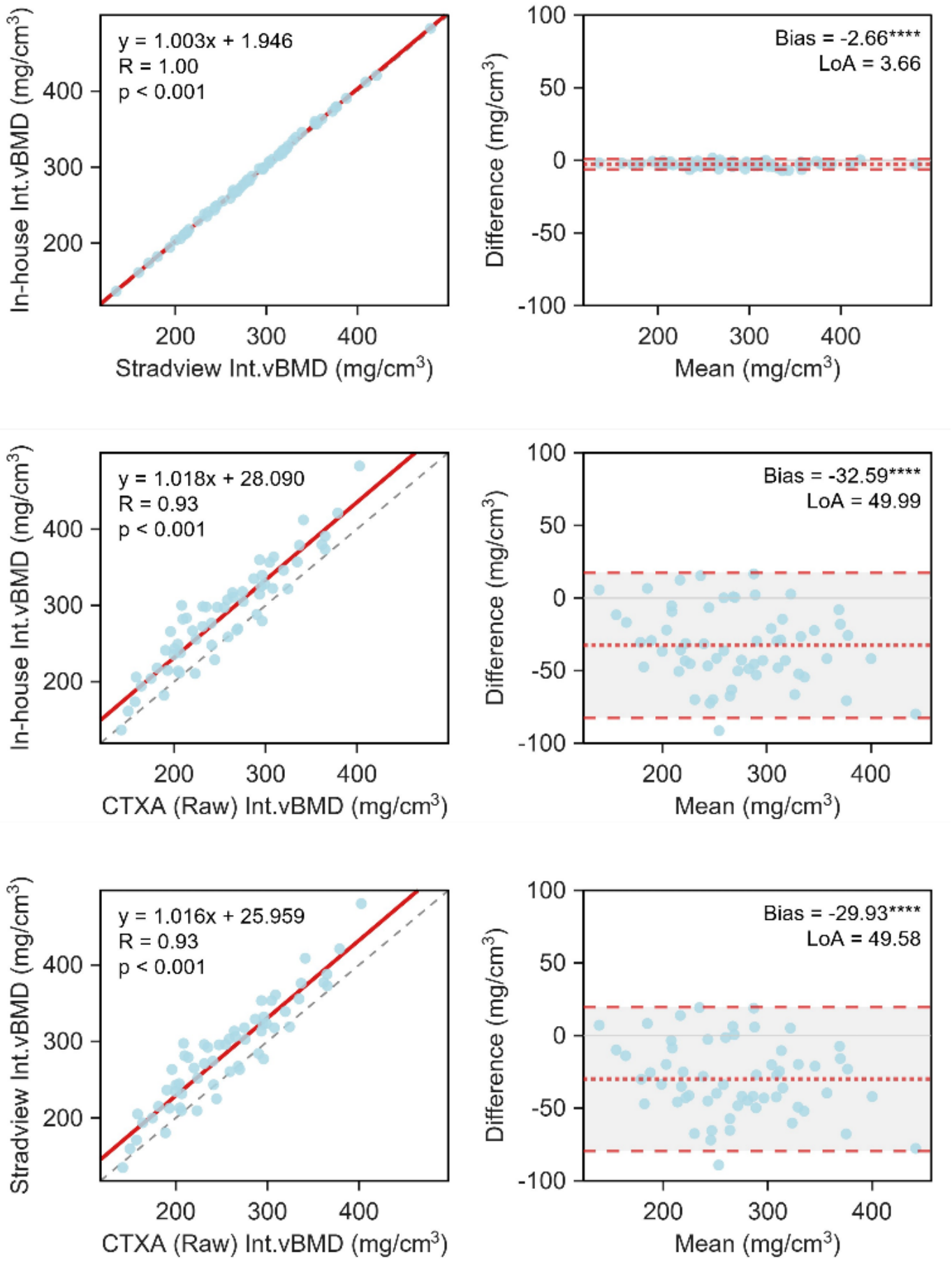

**Figure 2.** Agreement among QCT-based methods for integral vBMD, from top to bottom: *In-house Int.vBMD* vs *Stradview Int.vBMD*, *In-house Int.vBMD* vs *CTXA (Raw) Int.vBMD*, and *Stradview Int.vBMD* vs *CTXA (Raw) Int.vBMD*. Left column: scatter plots, with the solid red line showing the linear regression fit (equation, Pearson's R, and p-value given in each panel) and the dashed gray line showing the identity line (y = x) for reference. Right column: corresponding Bland-Altman plots, showing the difference between the two measurements (x-axis minus y-axis method of the scatter plot) against their mean; the dotted red line marks the mean bias, the dashed red lines mark the 95% LoA (bias ± 1.96 SD), and the shaded band spans the LoA. Asterisks indicate the significance of the bias (paired t-test): * p < 0.05, ** p < 0.01, *** p < 0.005, **** p < 0.001.

### 3.2.2. Trabecular vBMD

*In-house Tb.vBMD*, *Stradview Sc.vBMD*, and *Stradview Tb.vBMD* all agreed strongly with *3D-Shaper Tb.vBMD*, with regression slopes close to unity (0.957–0.992) and strong correlations (R = 0.92–0.94) (Figure 3). Bias and LoA were −11.7 mg/cm³ (LoA ±33.3 mg/cm³) for *In-house Tb.vBMD*, −7.8 mg/cm³ (LoA ±39.4 mg/cm³) for *Stradview Sc.vBMD*, and −11.8 mg/cm³ (LoA ±33.9 mg/cm³) for *Stradview Tb.vBMD. QCTPro CTXA* showed a different pattern: both *CTXA (Raw) Tb.vBMD* and *CTXA (Cor.) Tb.vBMD* correlated strongly with 3D-*Shaper Tb.vBMD* as well (R = 0.89 and R = 0.91), but with slopes well above unity (1.943 and 1.641), reflecting a proportional bias: differences were near zero at low trabecular density and became increasingly negative at higher density. Bias and LoA were −44.1 mg/cm³ (LoA ±61.3 mg/cm³) for *CTXA (Raw) Tb.vBMD* and −30.9 mg/cm³ (LoA ±54.3 mg/cm³) for *CTXA (Cor.) Tb.vBMD.*

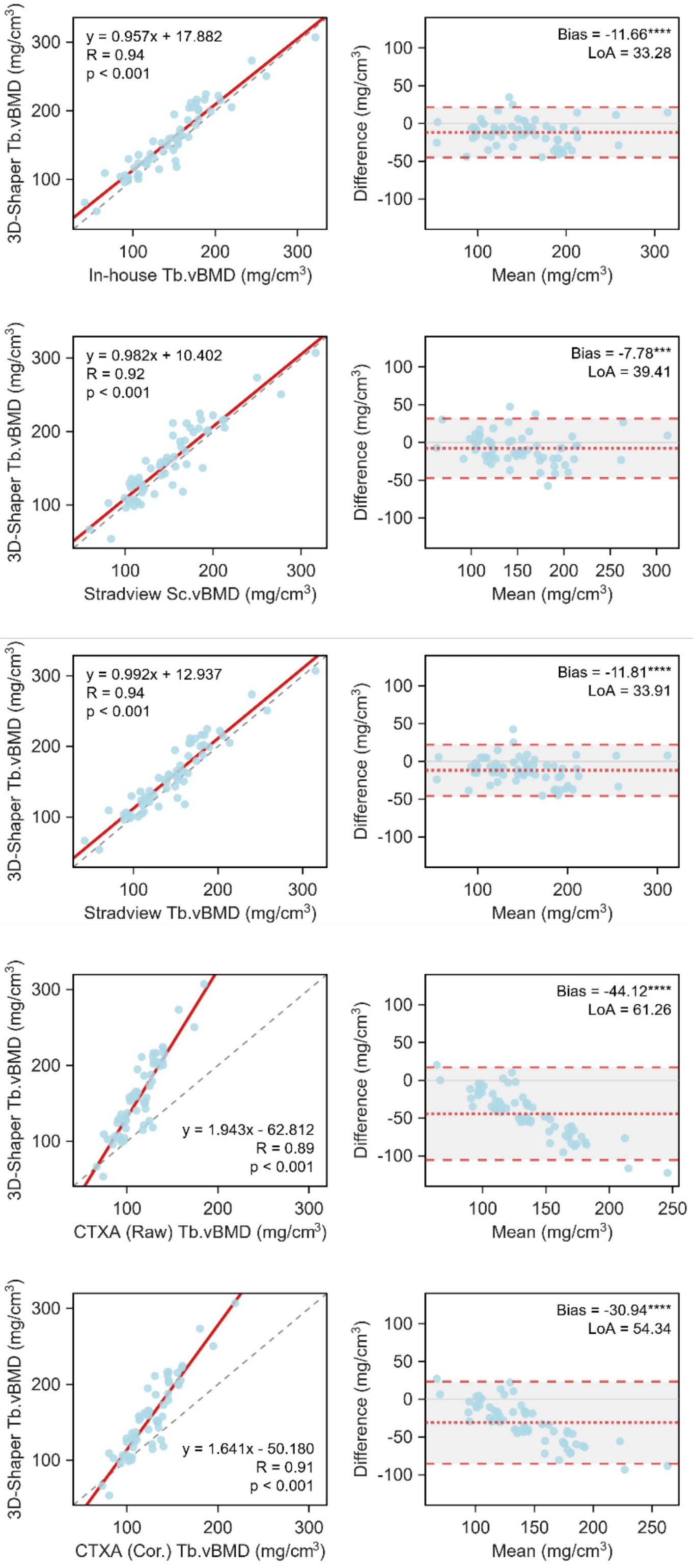

**Figure 3.** Agreement between *3D-Shaper Tb.vBMD* and QCT-based methods, from top to bottom: *In-house Tb.vBMD*, *Stradview Sc.vBMD*, *Stradview Tb.vBMD*, *CTXA (Raw) Tb.vBMD*, and *CTXA (Cor.) Tb.vBMD*. Left column: scatter plots, with the solid red line showing the linear regression fit (equation, Pearson's R, and p-value given in each panel) and the dashed gray line showing the identity line (y = x) for reference. Right

column: corresponding Bland-Altman plots, showing the difference between the two measurements (x-axis minus y-axis method of the scatter plot) against their mean; the dotted red line marks the mean bias, the dashed red lines mark the 95% LoA (bias ± 1.96 SD), and the shaded band spans the LoA. Asterisks indicate the significance of the bias (paired t-test): * $p < 0.05$, ** $p < 0.01$, *** $p < 0.005$, **** $p < 0.001$.

Between the QCT-based methods, *In-house Tb.vBMD* agreed strongly with *Stradview Sc.vBMD*, with a slope close to unity (1.011), a strong correlation (R = 0.96), a bias of 3.9 mg/cm$^3$ (LoA ±27.5 mg/cm$^3$), and no visible trend in the Bland-Altman differences (Figure 4). Agreement was near-perfect with *Stradview Tb.vBMD* (slope 1.038, R = 1.00), with a bias of −0.2 mg/cm$^3$ and narrow LoA (±6.9 mg/cm$^3$). Against *CTXA (Raw) Tb.vBMD*, however, both In-*house Tb.vBMD* and *Stradview Sc.vBMD* showed the same proportional bias seen against *3D-Shaper*: strong correlations (R = 0.96 and R = 0.94) but slopes well above unity (2.048 and 1.913), with differences near zero at low trabecular density and increasingly negative at higher density. Bias and LoA were −32.5 mg/cm$^3$ (LoA ±54.9 mg/cm$^3$) for *In-house Tb.vBMD* vs *CTXA (Raw) Tb.vBMD*, and −36.3 mg/cm$^3$ (LoA ±51.2 mg/cm$^3$) for *Stradview Sc.vBMD* vs *CTXA (Raw) Tb.vBMD*.

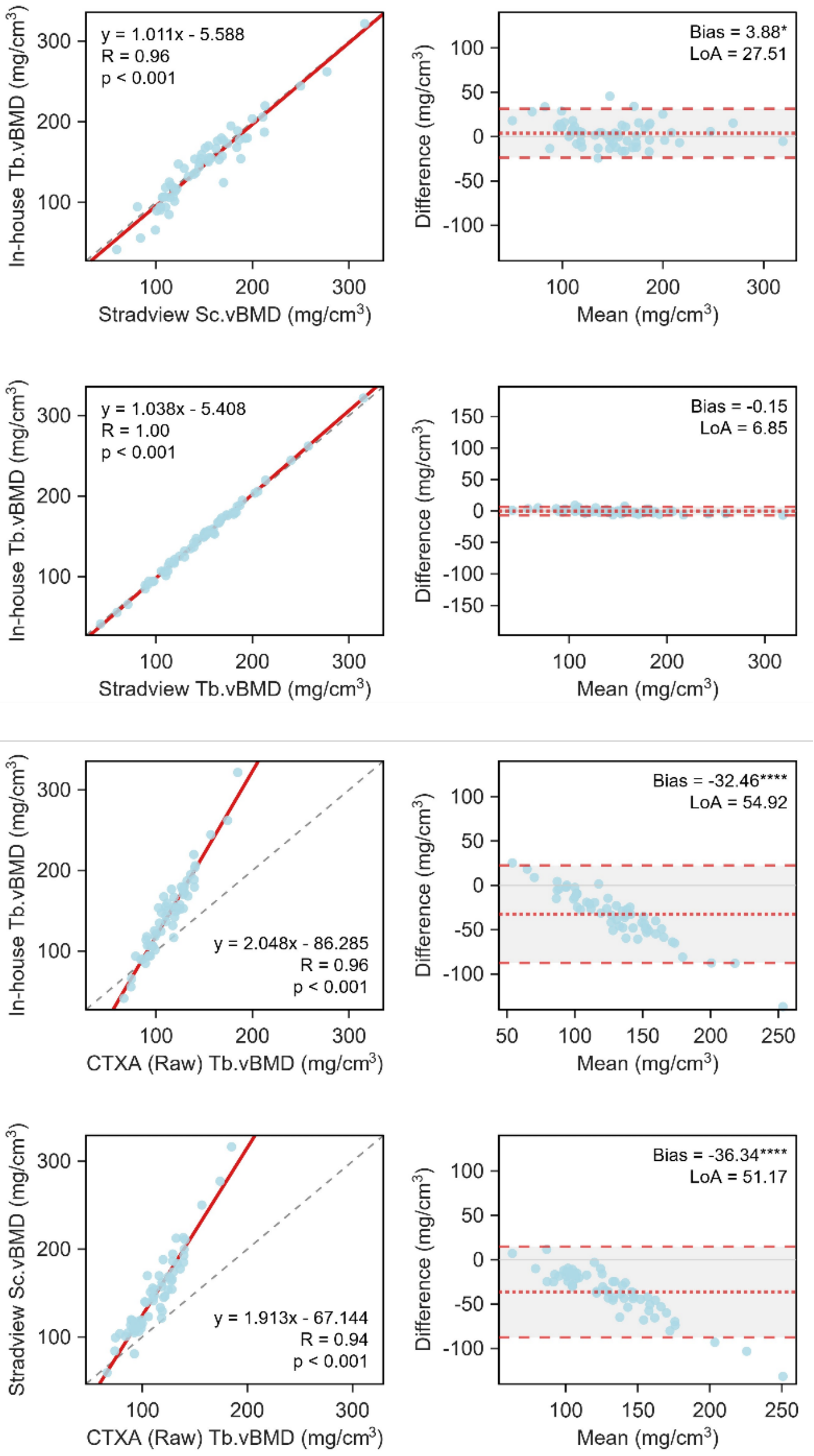

**Figure 4.** Agreement among QCT-based methods for trabecular vBMD, from top to bottom: *In-house Tb.vBMD* vs *Stradview Sc.vBMD*, *In-house Tb.vBMD* vs *Stradview Tb.vBMD*, *In-house Tb.vBMD* vs *CTXA (Raw) Tb.vBMD*, and *Stradview Sc.vBMD* vs *CTXA (Raw) Tb.vBMD*. Left column: scatter plots, with the solid red line showing the linear regression fit (equation, Pearson's R, and p-value given in each panel) and the dashed gray line showing the identity line (y = x) for reference. Right column: corresponding Bland-Altman plots, showing the difference between the two measurements (x-axis minus y-axis method of the scatter plot) against their mean; the dotted red line marks the mean bias, the dashed red lines mark the 95% LoA (bias ± 1.96 SD), and the shaded band spans the LoA. Asterisks indicate the significance of the bias (paired t-test): * $p < 0.05$, ** $p < 0.01$, *** $p < 0.005$, **** $p < 0.001$.

#### 3.2.3. Cortical sBMD

*In-house Ct.sBMD* agreed strongly with *3D-Shaper Ct.sBMD* (slope 0.817, R = 0.94), with a small bias of 5.6 mg/cm² (LoA ±23.6 mg/cm²) and no visible trend in the Bland-Altman differences (Figure 5). *Stradview Ct.sBMD* also correlated strongly with *3D-Shaper Ct.sBMD* (slope 0.762, R = 0.94), but with a substantially larger, consistent bias of 38.2 mg/cm² (LoA ±26.4 mg/cm²). This offset was largely preserved when *Stradview Ct.sBMD* was compared with *In-house Ct.sBMD* directly, which showed the strongest correlation of the three (slope 0.915, R = 0.98; bias 32.7 mg/cm², LoA ±14.6 mg/cm²). Unlike trabecular vBMD, *QCTPro CTXA* does not report cortical sBMD and is therefore not included in this comparison.

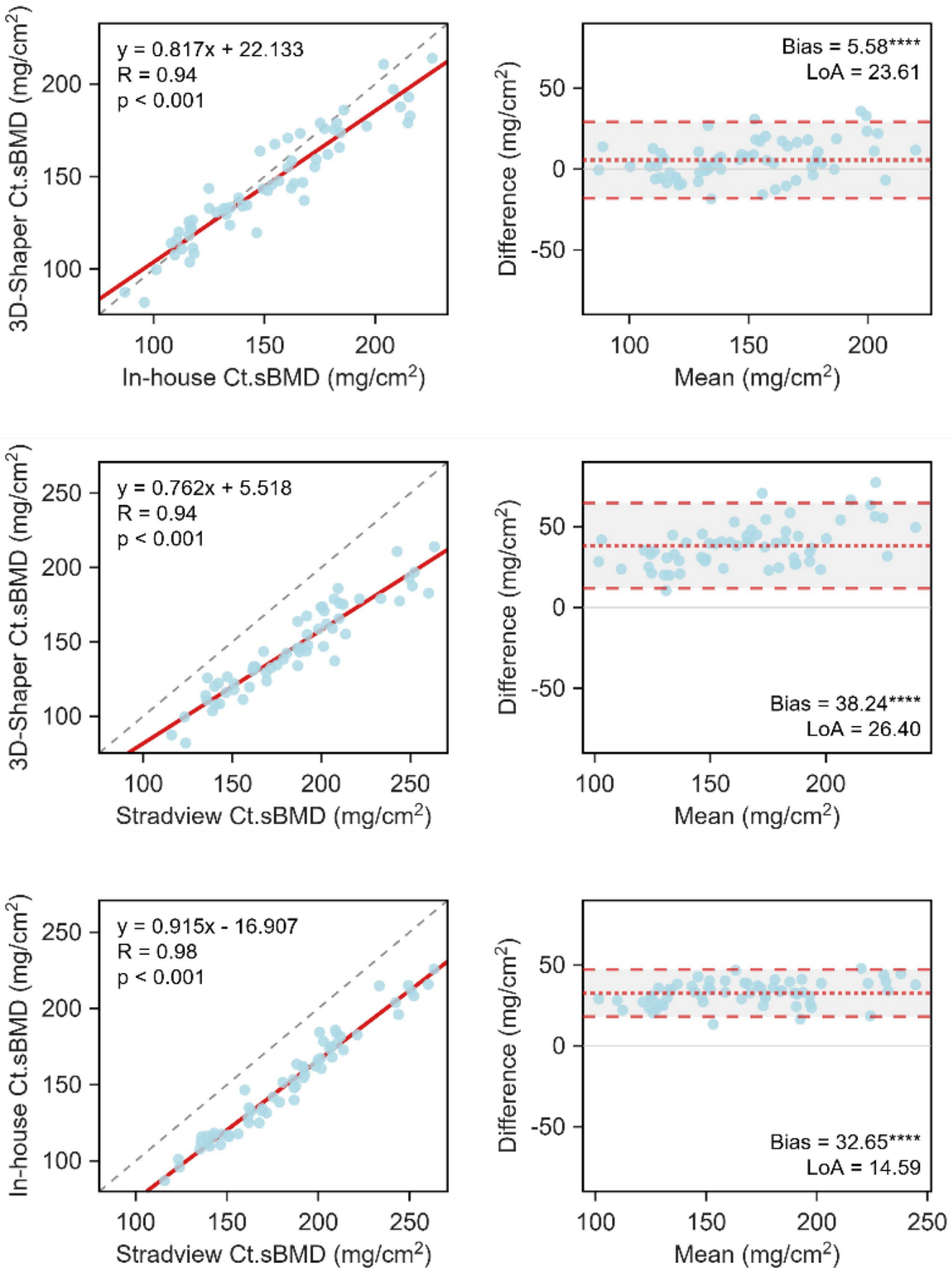

**Figure 5.** Agreement between methods for cortical sBMD, from top to bottom: *3D-Shaper Ct.sBMD* vs *In-house Ct.sBMD*, *3D-Shaper Ct.sBMD* vs *Stradview Ct.sBMD*, and *In-house Ct.sBMD* vs *Stradview Ct.sBMD*. Left column: scatter plots, with the solid red line showing the linear regression fit (equation, Pearson's R, and p-value given in each panel) and the dashed gray line showing the identity line (y = x) for reference.

Right column: corresponding Bland-Altman plots, showing the difference between the two measurements (x-axis minus y-axis method of the scatter plot) against their mean; the dotted red line marks the mean bias, the dashed red lines mark the 95% LoA (bias ± 1.96 SD), and the shaded band spans the LoA. Asterisks indicate the significance of the bias (paired t-test): * $p < 0.05$, ** $p < 0.01$, *** $p < 0.005$, **** $p < 0.001$.

## 4. Discussion

The present study evaluated the agreement of *3D-Shaper* with three independently implemented QCT post-processing methods for the assessment of proximal femur bone density and, importantly, also quantified the agreement among the QCT methods themselves. *3D-Shaper* showed close agreement with the in-house QCT pipeline and *Stradview* across integral vBMD, trabecular vBMD, and cortical sBMD, comparable to the agreement observed between these two QCT approaches. Agreement with *QCTPro CTXA* was more variable, particularly for trabecular vBMD, but the same pattern of disagreement was observed when *QCTPro CTXA* was compared directly with the other QCT methods. This provides important context for interpreting the validation of a DXA-based 3D reconstruction against QCT, because the QCT measurements themselves depend on the image processing approach used to define and quantify the cortical and trabecular compartments.

Agreement was strongest and most consistent for integral vBMD, which is intrinsically insensitive to the location of the cortical–trabecular boundary. This is consistent with Wang et al. [10], who also reported stronger agreement for integral vBMD than for compartment-specific measurements when comparing *QCTPro CTXA* against *MIAF-Femur*. In the present study, *3D-Shaper* showed strong correlations with the in-house pipeline and *Stradview* (R = 0.96 and 0.95), with biases of 4.2 and 1.5 mg/cm$^3$ and LoA of ±39.7 and ±41.4 mg/cm$^3$, respectively (Figure 1). Correlation with *QCTPro CTXA* was also strong (R = 0.90), although the mean bias was -28.4 mg/cm$^3$ and the LoA were wider (±61.5 mg/cm$^3$; Figure 1). Notably, integral vBMD measured with *QCTPro CTXA* showed similar biases when compared with the other QCT approaches: -32.6 mg/cm$^3$ versus the in-house pipeline (LoA ±50.0 mg/cm$^3$) and -29.9 mg/cm$^3$ versus *Stradview* (LoA ±49.6 mg/cm$^3$; Figure 2). For reference, a comparable absolute bias of 32.7 mg/cm$^3$ was reported by Wang et al. [10] for total femur integral vBMD between *QCTPro CTXA* and *MIAF-Femur* (R = 0.95). This consistency is noteworthy given the different cohorts and the different QCT software used as comparator (*MIAF-Femur* in Wang et al. [10] versus the in-house pipeline and *Stradview* here), and suggests that the observed offset in integral vBMD is related, at least in part, to methodological characteristics of *QCTPro CTXA* rather than being specific to its comparison with *3D-Shaper*. In the present study, PV correction reduced the mean *QCTPro CTXA* bias relative to *3D-Shaper* from -28.4 to -7.8 mg/cm$^3$ for integral vBMD (Figure 1), with a comparable reduction relative to the in-house pipeline (from -32.6 to -12.0 mg/cm$^3$; Supplementary Figure S1) and *Stradview* (from -29.9 to -9.3 mg/cm$^3$; Supplementary Figure S1), although in all three comparisons the LoA remained essentially unchanged.

For trabecular vBMD, *3D-Shaper* showed strong agreement with the in-house QCT pipeline and both *Stradview*-derived measurements, with slopes close to unity (0.957–0.992), R = 0.92–0.94, biases of −11.8 to −7.8 mg/cm$^3$ and LoA of ±33.3 to ±39.4 mg/cm$^3$ (Figure 3). Agreement between the in-house pipeline and *Stradview* was of similar or greater strength (R = 0.96 and 1.00, biases of 3.9 and −0.2 mg/cm$^3$, and LoA of ±27.5 and ±6.9 mg/cm$^3$ for *Stradview Sc.vBMD* and *Stradview Tb.vBMD*, respectively; Figure 4). *QCTPro CTXA*, in contrast, showed a clear proportional bias relative to *3D-Shaper*, the in-house pipeline and *Stradview Sc.vBMD*. For raw *QCTPro CTXA* trabecular vBMD, the slopes were 1.943, 2.048 and 1.913, the R values 0.89, 0.96 and 0.94, the biases −44.1, −32.5 and −36.3 mg/cm$^3$, and the LoA ±61.3, ±54.9 and ±51.2 mg/cm$^3$, respectively (Figures 3 and 4). PV correction reduced the slopes (1.641, 1.722 and 1.594), biases (−30.9, −19.3 and −23.2 mg/cm³) and LoA (±54.3, ±46.5 and ±44.4 mg/cm³), but did not remove the proportional bias (Figure 3 and Supplementary Figure S2). These LoA were broader than when *3D-Shaper*, the in-house pipeline and *Stradview* were compared with each other (±6.9 to ±39.4 mg/cm³), mainly as a consequence of the proportional bias, since the spread of the differences around the trend did not increase with density. These parallel patterns indicate that the proportional bias is not specific to the DXA-based reconstruction.

These findings are particularly relevant in light of the results previously reported by Wang et al. [10]. In their comparison of *QCTPro CTXA* and *MIAF-Femur*, trabecular BMD measured by *QCTPro CTXA* was 57 mg/cm³ higher on average at the total femur, with R = 0.94. More importantly, the bias was density-dependent: *QCTPro CTXA* was approximately 40 mg/cm³ higher at a mean trabecular BMD of 150 mg/cm³ and approximately 70 mg/cm³ higher at 80 mg/cm³ [10]. They attributed this behavior to the different treatment of the endocortical transition zone. This distinction matters given the limited spatial resolution of clinical QCT. At the proximal femur, cortical thickness can be below 1 mm, comparable to or smaller than typical QCT voxel dimensions, so PV effects at the endocortical boundary make separation of cortical and adjacent trabecular bone challenging [28]. With *QCTPro CTXA*, voxels in this transition zone are therefore classified according to whether their measured density falls above or below the fixed threshold. Because these voxels contain a mixture of cortical and trabecular signal due to PV effects, their contribution to the measured trabecular density depends on the underlying trabecular density; at higher trabecular densities, their contribution to the *QCTPro CTXA* trabecular value becomes relatively smaller. This mechanism could therefore account for the increasing negative differences observed here at higher trabecular densities (Figures 3 and 4). The present study extends the observation reported by Wang et al. [10] by showing a comparable proportional, density-dependent difference between *QCTPro CTXA* and three additional comparators: *3D-Shaper*, the in-house pipeline, and *Stradview*.

Cortical sBMD showed a strong relationship across *3D-Shaper*, the in-house pipeline, and *Stradview*. *3D-Shaper* correlated equally strongly with the in-house pipeline and *Stradview* (R = 0.94 for both; Figure 5). Although the bias between *3D-Shaper* and *Stradview* was larger (38.2 mg/cm²) than that between *3D-Shaper* and the in-house pipeline (5.6 mg/cm²), a similar offset was observed directly between *Stradview* and the in-house method (32.7 mg/cm²; Figure 5). This indicates that the offset is associated with differences between the model-based approaches rather than with *3D-Shaper* specifically. Cortical sBMD combines cortical thickness and cortical density into a surface-density measure. Because these quantities are difficult to resolve independently when the cortex is thin relative to imaging resolution, differences in their estimation can partially compensate when expressed as their product. This compensation is, however, only partial, because the in-house pipeline and *Stradview* resolve the thickness–density ambiguity in different ways. The in-house pipeline, like *3D-Shaper*, constrains the fit using a prior relationship between cortical thickness and density derived from micro-CT of cadaveric femurs [25], whereas *Stradview* first estimates a constant cortical density for the whole bone and then refines it locally according to the discrepancy between the locally fitted imaging blur and a global estimate of the scanner's blur [23]. This methodological difference may lead to systematic differences in absolute cortical sBMD values between the two methods, despite their very strong correlation (R = 0.98, Figure 5).

Another way to place these results in context is to compare the agreement between *3D-Shaper* and QCT with the agreement among the QCT methods themselves, since the latter defines the level of agreement that can reasonably be expected when the same bone is analyzed with different methods. The closest agreement was observed between the in-house pipeline and *Stradview*, with R = 0.96–1.00 across parameters and LoA of ±3.7 mg/cm³ for integral vBMD, ±6.9 to ±27.5 mg/cm³ for trabecular vBMD and ±14.6 mg/cm² for cortical sBMD (Figures 2, 4 and 5). This close agreement is expected given their similar working principles: both the in-house pipeline and *Stradview* separate the cortical and trabecular compartments by fitting a parametric model of the density profile across the cortex along lines normal to the femoral surface, accounting for the imaging blur, and differ mainly in how they resolve cortical thickness and density when the cortex is thin relative to the imaging resolution. *3D-Shaper*, which estimates these parameters from a single DXA projection, agreed somewhat less closely with both methods, with R = 0.92–0.96 and LoA of ±39.7 to ±41.4 mg/cm³ for integral vBMD, ±33.3 to ±39.4 mg/cm³ for trabecular vBMD and ±23.6 to ±26.4 mg/cm² for cortical sBMD (Figures 1, 3 and 5). However, these LoA were narrower than those observed when *QCTPro CTXA* was compared with the in-house pipeline and *Stradview* on the same CT scans, both for integral vBMD (±50.0 and ±49.6 mg/cm³) and for trabecular vBMD (±54.9 and ±51.2 mg/cm³; Figures 2 and 4). The same pattern was observed for the systematic differences. For integral vBMD, *3D-Shaper* showed no significant bias against the in-house pipeline or *Stradview* (4.2 and 1.5 mg/cm³; Figure 1), whereas *QCTPro CTXA* differed from both by −32.6 and −29.9 mg/cm³ (Figure 2). For trabecular vBMD, the biases between *QCTPro CTXA* and the other two QCT methods (−32.5 and −36.3 mg/cm³; Figure 4) were larger than those between *3D-Shaper* and the same methods (−11.8 to −7.8 mg/cm³; Figure 3), with slopes of 2.048 and 1.913 compared with 0.957–0.992 (Figures 3 and 4). These differences persisted after partial-volume correction, with LoA between corrected *QCTPro CTXA* trabecular vBMD and the in-house pipeline or *Stradview* (±46.5 and ±44.4 mg/cm$^3$; Supplementary Figure S2) remaining wider than those between *3D-Shaper* and the same two methods. In other words, for integral and trabecular vBMD, the differences between a DXA-based reconstruction and QCT were smaller than those between *QCTPro CTXA* and the other QCT methods. The differences between *3D-Shaper* and QCT therefore fall within the range of variability already observed among QCT methods, and should be interpreted in light of this variability rather than against an assumption of a single, method-independent QCT reference value.

These cross-method differences are also relevant when considering previous comparisons of *3D-Shaper* with QCT. In the STRUCTURE study, QCT measurements were obtained using *MIAF-Femur*, and the correlations with *3D-Shaper* (R = 0.84–0.87 for integral vBMD and R = 0.79–0.86 for trabecular vBMD [18]) were lower than those observed in the present study with the in-house pipeline and Stradview (R = 0.95–0.96 and R = 0.92–0.94, respectively). Given the method-dependent differences observed among QCT approaches in the present study, it would be informative to determine whether *MIAF-Femur* shows similarly different levels of agreement with the in-house pipeline and *Stradview*, which showed stronger agreement with *3D-Shaper* in the present study. Such comparisons could help distinguish differences attributable to the *3D-Shaper* reconstruction from those related to the QCT method used as comparator. Importantly, cortical sBMD could not be compared with *MIAF-Femur* in STRUCTURE, as *MIAF-Femur* does not calculate it, whereas in the present study it was evaluated against the other QCT methods and showed consistent agreement across the model-based approaches.

Several limitations should be considered. QCT acquisitions were performed using three scanner models across two centers, and scanner- and reconstruction-specific factors may have contributed to systematic differences between methods. Furthermore, images from the Philips scanner were reconstructed with the smoothest kernel (A) rather than the medium-smooth kernel (B), which increases imaging blur, and a thinner slice thickness relative to the slice spacing could have reduced partial-volume effects, although this affected all methods equally, since they were applied to the same images [28]. In addition, the *Stradview* trabecular and integral measurements (*Stradview Tb.vBMD* and *Stradview Int.vBMD*) used for direct comparison with *3D-Shaper* were derived from its exported periosteal and endocortical surfaces rather than from native global outputs, although this allowed definitions more closely aligned with those used for the other methods. Relatedly, *Stradview*'s CBM v2 algorithm is designed to provide local, vertex-level surface measurements rather than whole-bone averages, and the subject-level means used here for comparison with *3D-Shaper* may therefore not fully capture its intended measurement accuracy. Similarly, *QCTPro CTXA* is primarily designed to derive DXA-equivalent T-scores from CT scans, rather than to resolve compartment-specific cortical and trabecular geometry, which may partly explain its more variable agreement with *3D-Shaper* and the other QCT-based approaches for these measurements. However, the primary aim of this study was to evaluate *3D-Shaper* against multiple QCT approaches that independently define the cortical and trabecular compartments. Therefore, the comparisons were performed using measurements corresponding to the parameters provided by *3D-Shaper*, allowing the different methods to be evaluated within a common measurement framework. Moreover, the in-house QCT pipeline and *Stradview* used the region defined by *3D-Shaper*, matched to each QCT scan through registration with a technique that allowed the regions of interest to be aligned exactly, whereas *QCTPro CTXA* used its native regional definition, which is anatomically equivalent but was not computed in exactly the same way. This should therefore be considered when interpreting the absolute differences with *QCTPro CTXA*, although it does not readily account for the consistent methodological patterns observed across the different comparisons. Finally, although three QCT analysis approaches were evaluated, other QCT software solutions are available for proximal femur analysis, such as *MIAF-Femur*, which were not included in this comparison, and the agreement of *3D-Shaper* with these methods may differ.

A major strength of this study is that *3D-Shaper* was evaluated against multiple independently implemented QCT approaches rather than against a single reference method, as has been done before [11,22], while the agreement among the QCT approaches themselves was assessed in parallel. This makes it possible to distinguish differences attributable to *3D-Shaper* from those arising from the particular QCT method used as reference. Overall, *3D-Shaper* showed close agreement with the in-house QCT pipeline and *Stradview* and, for both integral and trabecular vBMD, showed smaller biases and narrower LoA with these two methods than *QCTPro CTXA* did, although correlations were similar.

## 5. Conclusion

In conclusion, *3D-Shaper* showed close agreement with the in-house QCT pipeline and *Stradview* for integral vBMD, trabecular vBMD, and cortical sBMD, with agreement across these parameters within the range observed among the QCT approaches themselves. Differences observed with *QCTPro CTXA*, particularly for trabecular vBMD, were also present when *QCTPro CTXA* was compared with the other QCT methods, indicating that these discrepancies are not specific to the DXA-based reconstruction. Overall, these findings support the use of *3D-Shaper* as a clinically accessible 3D-DXA approach for assessing integral vBMD, trabecular vBMD, and cortical sBMD at the proximal femur, while highlighting the importance of considering the analytical method used for QCT when interpreting cross-method comparisons.

## 6. Acknowledgments

The authors thank Professor Graham Treece (University of Cambridge), developer of *Stradview*, for his technical assistance with the software and its application in this study, and for his informal review of the original manuscript.

## 7. Founding sources

This research did not receive any specific grant from funding agencies in the public, commercial, or not-for-profit sectors.

## 8. Declaration of Interest

A.M. is an employee of 3D-Shaper Medical. L.H. is an employee and shareholder of 3D-Shaper Medical. The remaining authors declare no competing interests.

## Supplementary Material

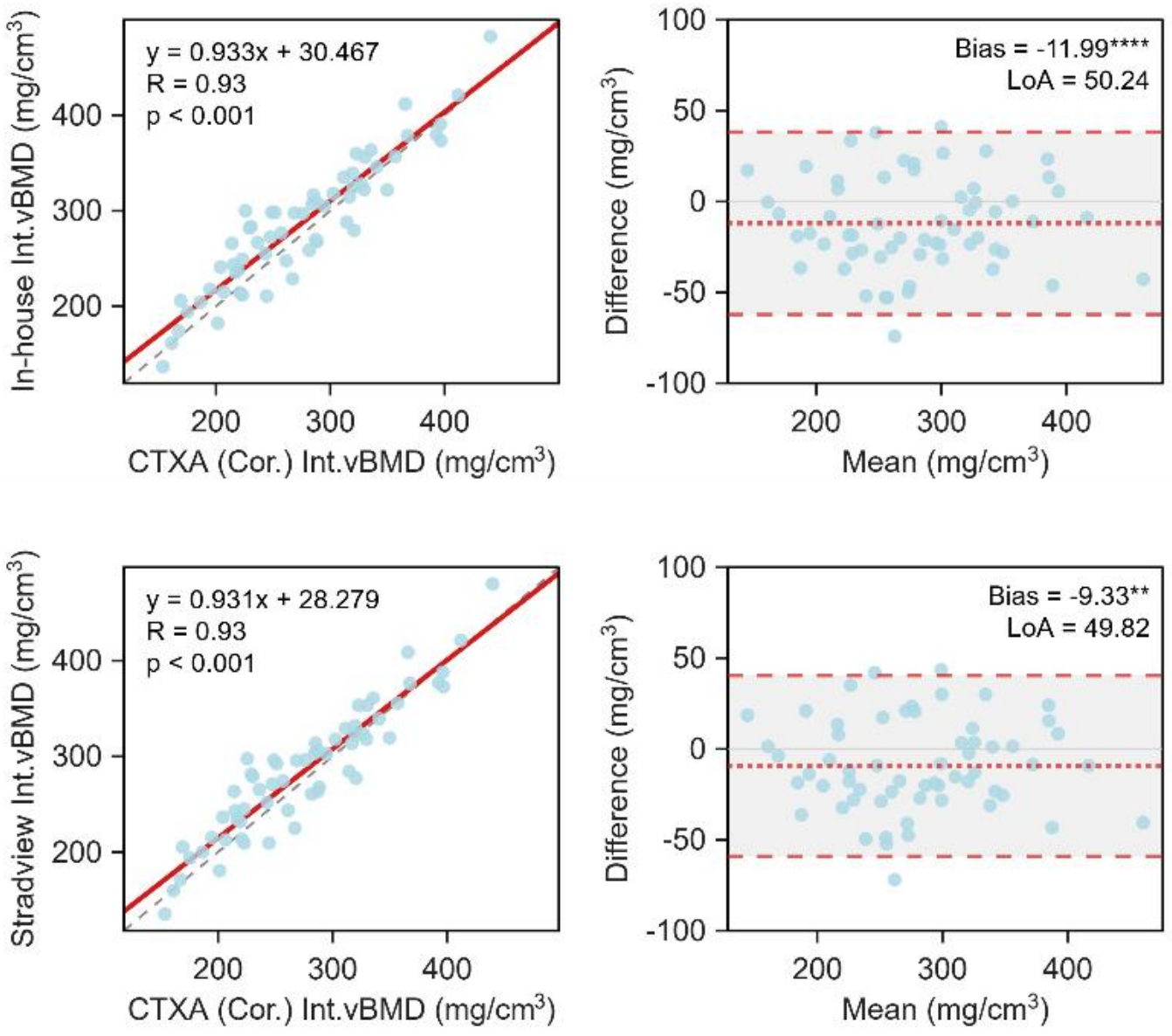

**Supplementary Figure S1.** Additional comparisons between *QCTPro CTXA* and the other QCT-based methods for integral vBMD, from top to bottom: *In-house Int.vBMD* vs *CTXA (Cor.) Int.vBMD*, and *Stradview Int.vBMD* vs *CTXA (Cor.) Int.vBMD*. Left column: scatter plots, with the solid red line showing the linear regression fit (equation, Pearson's R, and p-value given in each panel) and the dashed gray line showing the identity line (y = x) for reference. Right column: corresponding Bland-Altman plots, showing the difference between the two measurements (x-axis minus y-axis method of the scatter plot) against their mean; the dotted red line marks the mean bias, the dashed red lines mark the 95% LoA (bias ± 1.96 SD), and the shaded band spans the LoA. Asterisks indicate the significance of the bias (paired t-test): * $p < 0.05$, ** $p < 0.01$, *** $p < 0.005$, **** $p < 0.001$.

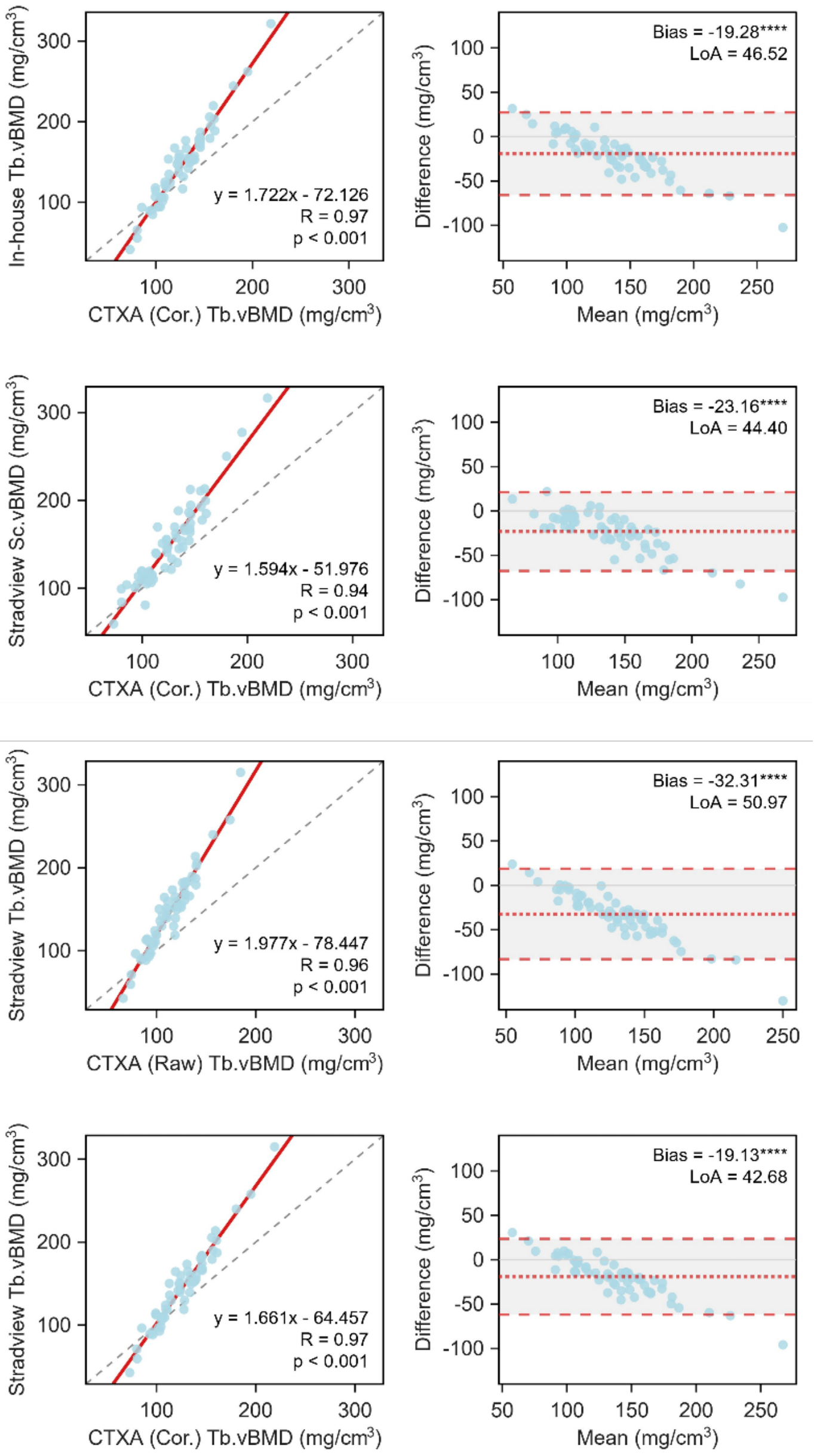

**Supplementary Figure S2.** Additional comparisons between *QCTPro CTXA* and the other QCT-based methods for trabecular vBMD, from top to bottom: *In-house Tb.vBMD* vs *CTXA (Cor.) Tb.vBMD*, *Stradview Sc.vBMD* vs *CTXA (Cor.) Tb.vBMD*, *Stradview Tb.vBMD* vs *CTXA (Raw) Tb.vBMD*, and *Stradview Tb.vBMD* vs *CTXA (Cor.) Tb.vBMD*. Left column: scatter plots, with the solid red line showing the linear regression fit (equation, Pearson's R, and p-value given in each panel) and the dashed gray line showing the identity line ($y = x$) for reference. Right column: corresponding Bland-Altman plots, showing the difference between the two measurements (x-axis minus y-axis method of the scatter plot) against their mean; the dotted red line marks the mean bias, the dashed red lines mark the 95% LoA (bias ± 1.96 SD), and the shaded band spans the LoA. Asterisks indicate the significance of the bias (paired t-test): * $p < 0.05$, ** $p < 0.01$, *** $p < 0.005$, **** $p < 0.001$.